\documentclass[%
preprint,
 amsmath,amssymb,
 aps,
]{revtex4-2}

\usepackage{graphicx}
\usepackage{dcolumn}
\usepackage{bm}

\RequirePackage[normalem]{ulem}
\RequirePackage{color}\definecolor{RED}{rgb}{1,0,0}\definecolor{BLUE}{rgb}{0,0,1}\definecolor{GREEN}{rgb}{0,1,0}
\providecommand{\ORadd}[1]{{\protect\color{red}\uwave{#1}}}
\providecommand{\ORrm}[1]{{\protect\color{red}\sout{#1}}}

\begin{document}

\title{Benchmarking exchange-correlation potentials with the mstar60 dataset: Importance of the nonlocal exchange potential for effective mass calculations in semiconductors}

\author{Magdalena Laurien}
 \email{laurienm@mcmaster.ca}

\author{Oleg Rubel}%
 \email{rubelo@mcmaster.ca}
\affiliation{%
 Department of Materials Science and Engineering, McMaster University, 1280 Main St W, Hamilton ON L8S 4L8, Canada
}%

\date{\today}

\begin{abstract}
The accuracy of effective masses predicted by density functional theory depends on the exchange-correlation functional employed, with nonlocal hybrid functionals giving more accurate results than semilocal functionals. In this article, we benchmark the performance of the Perdew–Burke–Ernzerhof (PBE), Tran-Blaha modified Becke-Johnson (TB-mBJ), and the hybrid Heyd-Scuseria-Ernzerhof (HSE06) exchange-correlation functionals and potentials for the calculation of effective masses with perturbation theory. We introduce the \texttt{mstar60} dataset, which contains 60 effective masses derived from 18 semiconductors. The ratio between experimental and calculated effective masses is $1.70 \pm 0.20$ for PBE, $0.76 \pm 0.04$ for TB-mBJ, $0.99 \pm 0.04$ for HSE06. 
We reveal that the nonlocal exchange in HSE06 enlarges the optical transition matrix elements leading to the superior accuracy of the hybrid functional in the calculation of effective masses. The omission of nonlocal exchange in the transition operator for HSE leads to serious errors. For the semilocal PBE functional, the errors in the bandgap and the optical transition matrix elements partially cancel out in the calculation of effective masses. The TB-mBJ functional yields PBE-like matrix elements paired with realistic bandgaps leading to a consistent overestimation of effective masses. However, if only limited computational resources are available, experimental masses can be estimated by multiplying TB-mBJ masses with the factor of $0.76$. We then compare effective masses of transition metal dichalcogenide bulk and monolayer materials: we show that changes in the matrix elements are important in understanding the layer-dependent effective mass renormalization.

\end{abstract}

\maketitle


\section{Introduction}
The effective mass is an important parameter in materials design and selection. It serves as an indicator of the carrier mobility, conductivity and the thermoelectric figure of merit and is often included in high-throughput computational material studies ~\citep{hautier_how_2014,curtarolo_aflowliborg_2012,hautier_identification_2013,tsymbalov_machine_2021,haastrup_computational_2018}.
The effective mass can be obtained from experimental measurements such as cyclotron resonance, Shubnikov-de-Haas oscillations and angle-resolved photoemission spectroscopy (ARPES).

The effective mass $m^*$ is inversely proportional to the energy band dispersion.
In the nearly-free electron model, the energy dispersion of a free electron is described by a parabola: $E = \hbar^2 k^2 / (2 m_0)$. In crystalline materials, the electron is no longer free as it interacts with the periodic potential of the ionic lattice. To describe the energy dispersion of the nearly free electron near a band maximum or minimum of interest in crystalline materials, particularly semiconductors, the mass of the electron $m_0$ in the parabola is replaced by an effective mass $m^*$ that acts as a scaling term to adjust the band curvature. 

The standard procedure for theoretically calculating the effective mass is to fit the band of interest with a parabola and obtain the effective mass from the curvature. An elegant alternative is to use perturbation theory.
From perturbation theory, we calculate the inverse effective mass $(m_{\alpha \beta, n}^{*})^{-1}$ for non-degenerate bands at a certain $k$-point as the following ~\cite[Appendix E]{ashcroft_solid_1976}
\begin{equation}\label{Eq:mstar}
\frac{m_{0}}{m_{\alpha \beta, n}^{*}}=\delta_{\alpha \beta}+\frac{1}{m_{0}} \sum_{l \neq n} \frac{p_{n l}^{(\alpha)} p_{l n}^{(\beta)}+p_{n l}^{(\beta)} p_{l n}^{(\alpha)}}{E_{n}-E_{l}},
\end{equation}
where $m_{0}$ is the electron rest mass, $\alpha$ and $\beta$ indicate directions in Cartesian coordinates $(x,y,z)$, $\delta_{\alpha \beta}$ is the Kronecker delta, the summation is over the band index $l$ but excludes the band of interest $n$. $E_{n}$ and $E_{l}$ denote the band energies and $p_{l n}$ the optical transition matrix element. The $k$ point index is omitted for simplicity. 

This equation helps us to develop some intuition about the factors influencing the effective mass: The larger the interband energy difference term $E_{n}-E_{l}$, the less the interaction between the bands contributes to the band dispersion. As a result, the larger the band gap the heavier is normally the effective mass. The larger the matrix element term (the numerator of the sum), the larger the contribution to the band curvature will be. Also, the wavefunctions of two bands can only couple if symmetry selection rules are fulfilled. Otherwise, the transition is not allowed and the matrix element $p_{ln}$ is zero~\citep[chap.~2.6.1]{yu_fundamentals_2010}. For interatomic transitions, i.e. from cation to anion, the effect of the matrix element on the band curvature can be understood in the tight-binding framework~\cite{kim_towards_2010,hautier_how_2014,zeier_thinking_2016} and \cite[chap. 2.7]{yu_fundamentals_2010}. The squared matrix element describes the probability of the transition and is thus related to the two-center hopping or overlap integral of tight-binding theory. Increased overlap between neighbouring orbitals leads to greater band dispersion. 
We should also note that bands lower in energy than $n$ make a positive contribution to the band curvature, while bands higher in energy than $n$ make a negative contribution due to the negative energy difference~\citep[chap.~2.6.1]{yu_fundamentals_2010}. 

The effective mass of charge carriers can be predicted with density functional theory (DFT)~\cite{Kohn_PR_140_1965} which is a ground-state theory. Still, DFT is commonly used to calculate excited state properties. Density functional approximations using semilocal exchange-correlation (XC) energy functionals such as the local density or generalized gradient approximation (LDA or GGA) are known to underestimate the band gap of semiconductors significantly~\cite[chaps.~6.3.1, 9.2.4]{bechstedt_many-body_2015}. This leads to errors in the band curvature and effective masses. Corrections from many-body theory change the band dispersion~\cite[chap.~16.1.3]{bechstedt_many-body_2015} and thus also cause an effective mass renormalization for many materials. 

The most obvious renormalization after the band gap correction comes from a relative change in the interband energy difference term $E_{n}-E_{l}$ in Eq.~(\ref{Eq:mstar}). This has recently been illustrated for InSe in a comparison of LDA and $GW$ calculations, where \citet{li_many-body_2020} showed that the out-of-plane electron effective mass was corrected three times more strongly than the in-plane mass as a result of the band gap correction in GW. This effect was explained by symmetry selection rules that ruled out a transition matrix element $p_{cv}$ between conduction and the valence band edge for the in-plane mass thus engaging deeper valence states ($p_{c,v-1}$) whose energy position relative to the conduction band edge is less affected by the correction of the fundamental band gap. However, what remains overlooked in Ref.~\citenum{li_many-body_2020} is that not only the band gap but also the $p_{ln}$ matrix elements are renormalized as we transition from LDA to a higher level of theory. The latter will be a central topic of this paper.

To predict more accurate effective masses with DFT, we should first find ways to correct the band gap inexpensively. This can be done using the semilocal Tran-Blaha modified Becke-Johnson exchange-correlation potential (TB-mBJ)~\cite{becke_simple_2006,tran_accurate_2009}. Interestingly, effective masses obtained with TB-mBJ are consistently heavier than the experimental result~\cite{kim_towards_2010,dixit_electronic_2012,araujo_electronic_2013,rubel_perturbation_2021}. The more expensive hybrid functionals on the other hand result in excellent agreement of effective masses with experiment~\cite{kim_towards_2010}. \citet{kim_towards_2010} alluded that to obtain accurate effective masses, corrections beyond a semilocal potential will ultimately be required. 
If the band gap is almost correct in TB-mBJ, then the transition matrix element must be underestimated. The role of the matrix element $p_{ln}$ for the renormalization of calculated effective masses has not yet been investigated in detail.

In this work, we benchmark the accuracy of effective masses calculated with several exchange-correlation potentials for a new dataset that we call \texttt{mstar60}. Our dataset comprises standard sp-semiconductors, d-semiconductors and monolayer materials. Effective masses are calculated with a perturbation theory approach. We show the extent of renormalization of effective masses caused by changes in the transition matrix elements. We explain the role of the nonlocal exchange potential $V_x^{NL}$ concerning these renormalization effects. On average 30\% heavier masses are predicted with the hybrid functional if incorrect transition matrix elements---that do not include $V_x^{NL}$---are used.

\section{Optical transition matrix elements \protect}

For the calculation of optical properties, the nonlocality of the potential becomes important when the transition matrix elements are calculated.
The matrix elements can be evaluated in the velocity gauge or the length gauge~\cite[chap.~20.1.1]{bechstedt_many-body_2015}. Assuming the dipole approximation, the coupling of electrons with an external electromagnetic field is described by $\mathbf{E \cdot r}$ in the length gauge and $\mathbf{A \cdot p}$ in the velocity gauge~\cite[chap. 5-1]{bassani_electronic_1975}\cite{marti_general_2005,del_sole_optical_1993}. Charge conservation and gauge invariance require the equivalence of the two interaction terms~\cite{del_sole_optical_1993,adolph_nonlocality_1996}.  

In the length or longitudinal gauge, the position operator $\boldsymbol{r}$ is used for the calculation of the optical transition matrix elements (in atomic units)~\cite{rhim_fully_2005}
\begin{equation}\label{Eq:length}
p_{n l}=\lim _{q \rightarrow 0} q^{-1}(E_{l, \mathbf{k}+\mathbf{q}} - E_{n, \mathbf{k}})\left\langle\psi_{l, \mathbf{k}+\mathbf{q}}\left|e^{i \mathbf{q} \cdot \mathbf{r}} \right| \psi_{n, \mathbf{k}}\right\rangle,
\end{equation}
where $\psi_{l, \mathbf{k}}$ is the single-particle wavefunction and $\mathbf{q}$ is a small momentum vector shift. 

In the velocity gauge (also called transverse or Coulomb gauge) transition matrix elements are calculated from the velocity operator (in atomic units) $\hat{v}$~\cite{starace_length_1971}:
\begin{equation}\label{Eq:velocity}
p_{n l}= \langle \psi_{l, \mathbf{k}}|\hat{v}| \psi_{n, \mathbf{k}}\rangle.
\end{equation}
The velocity operator is expressed as the commutator of the Hamiltoninan and the position operator $\hat{v}(\boldsymbol{r})=i[H,\boldsymbol{r}]=\hat{p}+i[V^{\mathrm{NL}}(\boldsymbol{r},\boldsymbol{r}'),\boldsymbol{r}]$. For local potentials $V(\boldsymbol{r})$, the velocity operator $\hat{v}$ is equivalent to the momentum operator $\hat{p}$. Therefore, in many cases the velocity matrix element $\langle \psi_{l, \mathbf{k}}|\hat{v}| \psi_{n, \mathbf{k}}\rangle$ is substituted by the momentum matrix element $\langle \psi_{l, \mathbf{k}}|\hat{p}| \psi_{n, \mathbf{k}}\rangle$ in the velocity gauge. However, for nonlocal potentials $V^{\mathrm{NL}}(\boldsymbol{r},\boldsymbol{r}')$ the position operator no longer commutes with the nonlocal potential and the velocity operator is no longer equivalent to the momentum operator. As a result, in order to calculate transition matrix elements from nonlocal potentials in the velocity gauge, the velocity operator has to be used or else a nonlocal correction to the momentum operator needs to be applied ($\hat{p}+i[V^{\mathrm{NL}}(\boldsymbol{r},\boldsymbol{r}'),\boldsymbol{r}]$). Otherwise the gauge invariance is violated~\cite{del_sole_optical_1993,adolph_nonlocality_1996, starace_length_1971}. In other words, for accurate optical matrix elements, the nonlocal potential must not be neglected. In the length gauge, nonlocal potentials are treated correctly automatically.

The nonlocality in the potential stems from the fact that the full-electron Hamiltonian is replaced by an approximate Hamiltonian in the independent-electron approximation with an effective potential that reintroduces electron-electron interactions in the Kohn-Sham equations~\cite[chap. 2]{giustino_materials_2014}. There are several sources by which nonlocality may be introduced in the effective Hamiltonian~\cite{adolph_nonlocality_1996, bechstedt_many-body_2015, pickard_second-order_2000}: an incomplete basis set, local field effects due to abrupt changes in the charge density (spatial inhomogeneity), nonlocal pseudopotentials and nonlocal exchange-correlation potentials or quasiparticle self-energies.

The importance of using the nonlocal correction in the velocity gauge has been widely discussed for nonlocal pseudopotentials~\citep{ismail-beigi_coupling_2001, pickard_second-order_2000,gajdos_linear_2006}. It was shown that neglecting the nonlocal term in the velocity gauge leads to inaccurate matrix elements, especially for transitions that involve localized d-electrons~\cite{read_calculation_1991}.  Also, several works have investigated the nonlocal effects of the self-energy operator on transition matrix elements from many-body $GW$ calculations~\cite{levine_optical_1991,adolph_nonlocality_1996,del_sole_optical_1993}. At the DFT level, ~\citet{rhim_fully_2005} calculated optical matrix elements including non-local exchange with the screened-exchange LDA functional (sX-LDA). They showed that to obtain the correct band dispersion, opening the bandgap with a scissor operator is not enough and the full calculation of the matrix element effects is necessary. Further, \citet{paier_dielectric_2008} showed that including nonlocal exchange via hybrid functionals yields more accurate static and dynamic dielectric functions in comparison with semilocal functionals. 
In this work, we focus on the nonlocality introduced by a nonlocal hybrid exchange-correlation potential and its effect on the accuracy of calculated effective masses.

\section{Methods}
\subsection{Dataset}
The data set contains 14 bulk and 4 monolayer materials with a total of 60 effective masses. Materials considered include sp-semiconductors and d-element semiconductors containing one transition metal. We include both three-dimensional and two-dimensional (layered) structures. The materials considered cover a wide range of effective masses. For the creation of the effective mass data set, we collected experimental effective mass data from the available literature. Most experimental data was taken from existing compilations in the Landolt-B{\"o}rnstein database~\cite{noauthor_landolt-bornstein_nodate}. We also included several individual entries from the literature for layered and monolayer 2D materials. Wherever multiple experimental values of one effective mass were available, we took the average for comparison with our computational data. For the compilation of the database, we had to exclude materials for which the experimentally reported effective masses differed widely, as this rendered comparison with computational results unprofitable.

\subsection{DFT functionals}
To benchmark our dataset, we computed the effective masses using three different exchange-correlation potentials and compared the calculated effective masses to experimental data. The potentials represent different levels of theory, and were chosen with the purpose of estimating how the effective masses are affected by the nonlocal potential which is represented differently with each functional.

The first level of theory was the Perdew–Burke–Ernzerhof (PBE)~\cite{perdew_generalized_1996} GGA exchange-correlation functional, which is semilocal in its treatment of exchange and correlation and thus does not include a nonlocal exchange potential. 

We also computed the effective masses with the Tran-Blaha modified Becke-Johnson potential (TB-mBJ)~\cite{becke_simple_2006,tran_accurate_2009} potential which corresponds to the second level of theory. The TB-mBJ potential 
\begin{equation}
V_{x, \sigma}^{\mathrm{mBJ}}(\mathbf{r})=c V_{x, \sigma}^{\mathrm{BR}}(\mathbf{r})+(3 c-2) \frac{1}{\pi} \sqrt{\frac{5}{6}} \sqrt{\frac{ t_{\sigma}(\mathbf{r})}{\rho_{\sigma}(\mathbf{r})}}
\end{equation}
is also a semilocal approximation. It is based on the Becke-Roussel~\cite{ becke_exchange_1989} potential $V_{x, \sigma}^{\mathrm{BR}}(\mathbf{r})$ which models the Coulomb potential of the exchange hole. $\sigma$ denotes the spin. Besides $V_{x, \sigma}^{\mathrm{BR}}(\mathbf{r})$, the TB-mBJ potential includes a term proportional to $\sqrt{ t_{\sigma}(\mathbf{r})/\rho_{\sigma}(\mathbf{r})}$, where $t_{\sigma}$ is the kinetic energy density and $\rho_{\sigma}$ is the electron density. This root term can be interpreted as a screening term~\cite{koller_merits_2011}. While semilocal in its approach, TB-mBJ mimics nonlocal effects. The parameter $c$ can be determined self-consistently. To estimate the prediction power of TB-mBJ potential, we first determined the $c$ parameter self-consistently to calculate the effective masses of our dataset. 
Additionally, to estimate the effect of nonlocal contributions, we adjusted the $c$ parameter to reproduce the experimental band gap to eliminate an additional source of data scattering when comparing effective masses with experiment and hybrid functional calculations.
The band gap was fitted with a maximum error of less than 2\%.

On the third level of theory, we used the Heyd-Scuseria-Ernzerhof hybrid functional (HSE06)~\cite{krukau_influence_2006} to compute effective masses. In a hybrid functional, a percentage of nonlocal Hartree-Fock (HF) exchange is mixed with the local PBE exchange-correlation functional. For HSE, the exchange is divided into a short-range (SR) and a long-range (LR) contribution. Only for the short-range exchange, a part of the PBE exchange is replaced by the exact Hartree-Fock exchange. The long-range exchange is entirely taken from the PBE functional. The HSE funtional takes the form
\begin{equation}
E_{\mathrm{xc}}^{\mathrm{HSE}}= a E_{\mathrm{x}}^{\mathrm{HF}, \mathrm{SR}}(\omega)+(1-a) E_{\mathrm{x}}^{\mathrm{PBE}, \mathrm{SR}}(\omega) +E_{\mathrm{x}}^{\mathrm{PBE}, \mathrm{LR}}(\omega)+E_{\mathrm{c}}^{\mathrm{PBE}},
\end{equation}
where $\omega$ denotes the range separation between SR and LR. There are various hybrid functionals that are distinguished by their inverse screening length or range separation $\omega$: PBE0 ($\omega = 0$), HSE06 ($\omega = 0.2$~{\AA}$^{-1}$), HSE03 ($\omega = 0.3$~{\AA}$^{-1}$). The parameter $a$ specifies the share of SR Hartree-Fock exchange included. It is typically set to 0.25. 

Because of the inclusion of a part of the exact nonlocal exchange, nonlocal exchange effects are considered explicitly in hybrid functionals. Hartree-Fock exchange is unscreened. The mixing of the nonlocal Hartree-Fock exchange with the local PBE exchange-correlation amounts to an effective screening of the nonlocal exchange by the local exchange-correlation~\cite{becke_perspective_2014}, leading to a very good agreement with experiment for the electronic structure of semiconductors. Due to this artificial screening, the HSE approach can be seen as an approximation to the $GW$ approach~\cite[chap. 9.2]{bechstedt_many-body_2015}. $GW$ includes the dynamically screened exchange $W$ in a physically correct way. 

In our work, we first estimated the prediction power of the standard HSE06 ($a = 0.25$) potential by calculating the effective masses and comparing them to experimental data.
Then, we fitted $a$ for each material to reproduce the experimental band gap with less than 2\% error. This allowed us to directly compare HSE and TB-mBJ effective masses with the goal of analyzing nonlocal effects on the optical matrix element.
Effects of different values for the range separation (i.e. by using PBE0 and HSE03 functionals) are discussed in the results section. 
To exclude the nonlocal exchange potential during the optical matrix element calculation, we switched to the PBE potential during the optical calculations while using HSE wavefunctions (see supplementary information for the detailed workflow). As an alternative to using the PBE potential, HSE calculations can be performed with the fraction of the exact HF exchange set to zero for this step. The effective masses obtained from the two methods are identical with a difference of less than 1\%. From this we conclude that the imperfect model PBE exchange hole in HSE does not affect our results.

\subsection{Computational details}
Density functional calculations were performed with the Vienna \textit{ab initio} simulation package~\cite{kresse_efficient_1996,kresse_efficiency_1996} (VASP), which uses projector-augmented waves~\cite{blochl_projector_1994} as basis set, implemented by \citet{kresse_ultrasoft_1999}. The plane wave cutoffs were taken from the values recommended in the pseudopotentials distributed with VASP. The number of valence and semicore electrons included for each element was chosen according to the values recommended by the Materials Project database~\cite{jain_commentary_2013}. For molybdenum and tungsten we included additional semi-core states (14 valence and semi-core electrons in total). The Brillouin zone was sampled with $\Gamma$-centered $k$ grids. Table~S1 of the supplementary information lists the $k$ grids used for each material. 

Experimental lattice parameters were used for all bulk systems, allowing only atomic positions to relax using the PBE functional with a force convergence criterium of 0.001~eV/{\AA}. Experimental structure data were obtained from \citet{wyckoff_crystal_1963} unless otherwise specified in Table~\ref{table-1}. The monolayers were obtained by theoretical exfoliation from the corresponding bulk material. To avoid interactions between periodic images of a monolayer, we included more than 25~\AA~of vacuum in the out-of-plane direction. Subsequently, the monolayers were fully relaxed on the PBE level. 

All systems were treated as non-magnetic. Spin-orbit coupling was included in all calculations. Additional system-dependent calculation parameters are recorded in Table~S1 of the supplementary information. Table~S1 lists the experimental bandgaps (Refs.~\cite{madelung_semiconductors_2004,kam_detailed_1982,evans_determination_2008,huang_bandgap_2015,ugeda_giant_2014,zhu_exciton_2015,zhang_probing_2015}) that were used to fit the HSE and TB-mBJ band gaps, the fitting parameters and the number of bands included in the optical calculations.
We performed optical calculations in VASP to compute the transition matrix elements. In VASP, the longitudinal gauge (see Eq.~(\ref{Eq:length})) is implemented for the calculation of the transition matrix elements~\cite{gajdos_linear_2006}. In this gauge, nonlocal potentials are evaluated correctly.

Effective masses were calculated with the \texttt{mstar} code~\cite{rubel_perturbation_2021} which uses a perturbation theory approach based on Eq.~(\ref{Eq:mstar}) and its extension for degenerate states. The perturbation approach includes a sum over all bands and therefore many empty bands have to be included for accurate effective mass calculations. This is especially true for heavy effective masses and band edges that interact with high-energy orbitals.  For the optical calculations, we included empty bands of up to 7 Ry (96 eV) above the Fermi level to ensure an accurate calculation of the effective mass. 7 Ry suffices for most materials but not for all, as we will discuss later. In contrast to Fourier expansion methods for calculating effective masses (as implemented in the \texttt{BoltzTraP} code~\cite{madsen_boltztrap_2006}), the perturbation approach does not require a dense $k$ grid to accurately capture light effective masses. 

\subsection{Statistics}
For the statistical analysis, we determined the mean error (ME), mean absolute error (MAE), mean relative error (MRE) and the mean absolute relative error (MARE) of the computational effective masses with respect to experimental values. The standard deviation of the  error (STDE) and the relative error (STDRE) were also calculated. 
For benchmarking our dataset with regard to prediction power of effective masses, we plotted the calculated effective masses over the experimental effective masses on a log-log scale. We calculated the linear regression, holding the slope constant at unity. From the intercept of the linear regression, we obtained the scaling factor between calculated effective masses and experiment.  
Only materials for which data for all functionals was collected were included in the statistical analysis. Errors with a z-score of more than 3.5 were treated as outliers.

\section{Results and discussion}

Table \ref{table-1} shows the results for 60 effective masses of 14 bulk and 4 monolayer materials obtained with the PBE, HSE06 and TB-mBJ exchange-correlation potentials. To keep with the convention, the sign of the valence band effective masses is inverted, that is, a valence band curving downwards yields a positive effective mass. Negative values denote an upwards-bent valence band or a downwards-bent conduction band.

\subsection{Accuracy of the method}

First, we establish the accuracy of our method. We compare the perturbation theory results at the PBE level with band curvature fits for which the band of interest was fitted in an energy window of 25~meV (the thermal energy at room temperature) with a fourth-order polynomial and extracted the second order coefficient. For GaAs we compared m$_{\mathrm{n}}$, m$_{\mathrm{p,hh}}$, m$_{\mathrm{p,lh}}$ and m$_{\mathrm{p,so}}$ and found that perturbation theory results agreed within an error of 1\% with the band curvature fit. For Si, m$_{\mathrm{p,hh}}$, m$_{\mathrm{p,lh}}$ and m$_{\mathrm{p,so}}$ agreed within 2.5\% error. For 1L MoS$_2$ at the K point the band curvature yields an effective mass which is 15\% smaller than the perturbation theory result. The conduction band effective mass from the band curvature is 7\% larger than the perturbation theory result.

For some monolayer effective masses at $\Gamma$ (Table~\ref{table-1}), reliable PBE-derived effective masses could not be obtained with perturbation theory as the result differed by more than 30\% from the band curvature fit. For effective masses at the K point, the errors with respect to the band curvature were in the range of $5-16$\% which is significantly larger than for sp-semiconductors. This is due to challenges with representing d-states using perturbation theory with DFT pseudopotentials. All monolayers considered in our study are transition metal dichalcogenides for which the band edges are composed mainly of the d-orbitals of the transition metal. When it comes to the prediction of d-states, many high-energy bands are required to converge the perturbation sum. Two factors are important, the number of bands and the energy of the bands. Firstly, the results have to be converged carefully with respect to the number of bands included. The need to sum over many empty states in the perturbative expansion can elegantly be overcome with the Sternheimer approach~\cite{sternheimer_nuclear_1951}. Secondly, we have to determine the limits of the pseudopotential for predicting high-energy states accurately. Many pseudopotentials represent coupling to high-energy states incorrectly. It is possible to address this issue by including high-energy local orbitals (HELOs) to augment the basis set. This feature available in the Wien2k code~\cite{blaha_wien2k_2019,blaha_wien2k_2000} was shown to improve effective masses derived from the perturbative expansion~\cite{rubel_perturbation_2021}. Details about converging perturbation theory calculations for materials with d-states can be found in Section~B of the supplementary information (cited in this SI section:~\cite{mattsson_validating_2014,giustino_gw_2010,laflamme_janssen_precise_2016,laskowski_calculating_2014,ren_all-electron_2021}). 

Previous effective mass calculations at the PBE level agree well with our results. For example, for the conduction band effective masses of Silicon, our results agree well with the ones obtained by \citet{zhong_first_2016} and \citet{yu_first-principles_2008} (in brackets), respectively: m$_{\mathrm{n}, \parallel} = 0.943$ (0.950; 0.95) and m$_{\mathrm{n},\perp}=0.193$ (0.197; 0.19) (all effective masses in units of m$_0$).
For GaAs, our values agree well with the results reported by \citet{kim_towards_2010} (in brackets): m$_{\mathrm{p,so}}=0.107~(0.108)$, m$_{\mathrm{p,lh}}=0.034~(0.036)$, m$_{\mathrm{p,hh}}=0.324~(0.320)$ and m$_{\mathrm{n}}=0.028~(0.030)$. For monolayer MoS$_2$ our data at the K point show satisfactory agreement with the results of \citet{wang_many-body_2014}, \citet{kormanyos_k_2015} and \citet{wang_vaspkit_2021} (in brackets), respectively: m$_{\mathrm{p}} (K) = 0.603~(0.59;0.56;0.54)$ , m$_{\mathrm{n}} (K) = 0.402~(0.5;0.47;0.47)$. 
Overall, our perturbation theory results are accurate with respect to band curvature fits and agree very well with previously published data, especially for sp-semiconductors.

\subsection{Statistical analysis}

Monolayer materials and black phosphorus were excluded from the statistical analysis due to incomplete effective mass data as explained below. Furthermore, the following effective masses were excluded from the statistical analysis as outliers with a z-score above 3.5: GaAs $\mathrm{m}_{\mathrm{n}, \parallel}$ (X$_6$), CdS $\mathrm{m}_{\mathrm{p}, \parallel}$ ($\Gamma$, A exciton), and BN m$_\mathrm{p}$ ($\bar{\mathrm{K}}$). The outliers are marked with $\dagger$ in Table \ref{table-1}. In total, 42 effective masses of 12 materials were included in the statistical analysis.

For the statistical analysis of the dataset, we included only materials for which effective masses were obtained with all three functionals. In spite of being very effective for most solids, the TB-mBJ potential (aw well as its local version~\cite{rauch_local_2020} designed for materials with vacuum) was unable to open the band gap beyond PBE for monolayers of MoS$_2$, MoSe$_2$, WS$_2$, and WSe$_2$ as also noted by \citet{patra_efficient_2021} and further explained by \citet{tran_bandgap_2021}. As a result, we did not calculate masses in monolayers with TB-mBJ and excluded them from the statistical analysis. Black phosphorus was also excluded because representative effective masses could not be obtained at with the PBE functional. At the PBE level, the conduction band of black phosphorus is lower in energy than the valence band, leading to a metallic ground state with band inversion and band mixing around the band edges. This causes effective masses of inverted sign and magnitude in two directions. A proper band order is restored at a higher level of theory (HSE, TB-mBJ).

\subsection{Benchmarking the prediction power of PBE, TB-mBJ, and HSE06 with the \texttt{mstar60} dataset}

Now we turn to benchmarking the PBE, TB-mBJ and HSE06 functionals with the \texttt{mstar60} dataset.

Figure~\ref{overview} shows experimental effective masses versus calculated effective masses obtained with PBE, TB-mBJ and HSE06. The colors represent the main contribution to the orbital composition of the band. The figure also shows the linear regression fits and the scaling factors derived from them. The linear regression includes all data entries of the \texttt{mstar60} dataset included in the statistical analysis. The scaling factor is the coefficient of the experimental and calculated effective masses and thus gives an estimate of the effective mass prediction power of a particular functional. 
The effective mass ratio of Exp./PBE is $1.70 \pm 0.20$. The ratios of Exp./TB-mBJ and Exp./HSE06 are $0.76 \pm 0.04$ and $0.99 \pm 0.04$, respectively. Comparing the ratios, we see that HSE06 has the best prediction power, followed by TB-mBJ. Thus, if only limited computational resources are available, experimental masses can be estimated by multiplying TB-mBJ masses with the factor of $0.76$. 

The superiority of the hybrid functional is also seen in the summary statistics. Table \ref{stats-bulk} shows the summary statistics for the \texttt{mstar60} dataset. 
Effective masses calculated with the TB-mBJ potential show the largest errors, with a mean absolute relative error of 45\% and a mean relative error of 39\%. The positive values of mean error and mean relative error suggest that effective masses are in many cases overestimated. This is also reflected in the Exp./TB-mBJ effective mass coefficient being smaller than 1. 
Effective masses calculated with the PBE functional also show large errors, with a mean absolute relative error of 38\% and a mean relative error of $-$27\%. The negative values of mean error and mean relative error suggest that effective masses are in many cases underestimated. However, the scattering of the error is large as indicated by the standard deviation of the relative error of 40\%. 
Effective masses calculated with the HSE06 functional show the best agreement with experiment with a mean absolute relative error of 21\% which is about half that of masses obtained with the PBE and TB-mBJ funcitonals. The mean relative error of 5.1\% is much smaller than the mean absolute relative error for HSE06 which shows that there is no clear trend for over- or underestimation of the effective mass using HSE06.

The main conclusion of the benchmarking of the PBE, TB-mBJ, and HSE06 exchange-correlation potentials is that HSE06 gives by far the best agreement with experimental effective masses. On the other hand, PBE often yields lighter effective masses, while TB-mBJ generally overestimates them. These trends are in agreement with the results of \citet{kim_towards_2010}. 

\subsection{Analysis of nonlocal effects}

Having confirmed that HSE06 gives the best results, our objective is to comprehend with greater clarity the properties that make hybrid functionals so successful in reproducing experimental effective masses. As stated earlier, the main difference of the hybrid functionals with respect to semilocal functionals is the addition of a nonlocal component via the introduction of a fraction of HF exchange. Effective masses are influenced by the nonlocal component in two ways: Firstly, by adding nonlocal exchange, the band gap opens up, which increases the effective mass. Secondly, nonlocal effects influence the optical transition matrix elements $p_{ln}$. The opening of the band gap can be reproduced with the TB-mBJ potential, but the errors in the effective masses are still much higher than with HSE. The larger errors with the TB-mBJ potential despite the band gap correction suggest that the optical transition matrix element is the key to accurate effective masses with hybrid functionals.

Therefore, we want to investigate how the calculation of the optical matrix element is affected by the nonlocal exchange potential $V_x^{\mathrm{NL}}$ present in hybrid functionals. For that, we need to control the other factor, which is the size of the bandgap. We fitted the HSE06 and TB-mBJ functionals for each material to reproduce the experimental band gap with less than 2\% error. This allowed us to directly compare HSE and TB-mBJ effective masses and optical matrix elements. Overtuned functionals can lead to qualitatively incorrect band structures in some cases. To test for this error, we plotted the bandstructures of several materials, including indirect-gap materials and monolayer 2D materials which needed the largest tuning to reproduce the quasiparticle gap. We found that the tuning of the functionals did not affect the quality of the bandstructures. Caution must be exercised for bands close in energy (in the order of 10 meV) as we found that the tuning may cause a change of band ordering in some cases (e.g. the conduction band mininum of WS$_2$ monolayer). The band ordering (spin up/down) was not crucial for our results and thus did not affect them.

Table \ref{stats-bulk} lists the summary statistics for the fitted functionals next to the standard functionals (the individual effective mass data of the fitted functionals are given in Table S3 of the supplementary information). The gap fitting considerably improves the accuracy of the TB-mBJ derived masses, reducing by half the relative error and its standard deviation as well as the mean relative error. For the HSE06 functional, the magnitude of several errors is also reduced, most notably the mean absolute relative error which is reduced by half. Still, after correcting for band gap effects, the relative errors of effective masses obtained with the HSE06 functional are significantly smaller than that of TB-mBJ.

In order to get an idea about the effect of the nonlocal exchange potential $V_x^{\mathrm{NL}}$ on the optical matrix element, it is instructive to switch off the nonlocal exchange contributions in the optical calculation step. We achieve this by switching to the PBE potential when calculating the matrix elements while using HSE wavefunctions. As an alternative to using the PBE potential, the fraction of the exact HF exchange can be set to zero in this step. The effective masses obtained are identical with a difference of less than 1\%.
As a result of switching functionals, HSE eigenfunctions and band gaps are conserved and only the matrix element is calculated without nonlocal exchange effects. This allows us to decouple the band gap increase and the matrix element change that are both caused by the nonlocal exchange.

Switching off the nonlocal exchange contributions to the matrix elements leads to a systematic overestimation of the effective masses (see Table~\ref{stats-bulk}, individual effective masses are given in Table S3 of the supplementary information). With respect to the full HSE06 calculations, the error is on average 30\% when nonlocal effects are neglected in the calculation of the matrix element. This must be a consequence of the absolute matrix element being \textit{smaller} when calculated without the nonlocal exchange potential.

To discuss the nonlocal exchange contribution to the effective mass in more detail, we now consider the example of the conduction band effective mass $m_\mathrm{n}$ of GaAs at $\Gamma$. Often, interactions between many bands influence effective masses and these effects are difficult to trace. The conduction band of GaAs, however, is non-degenerate and at the Gamma point its interactions with other bands are well confined. Contributions to the effective mass are almost entirely determined by interactions between the conduction band and the three highest valence bands. This makes $m_\mathrm{n}$ of GaAs at $\Gamma$ a simple system and an excellent example to understand the principle behind the nonlocal exchange effect. 

\begin{figure}[h] 
	\includegraphics[scale=1, angle=0]{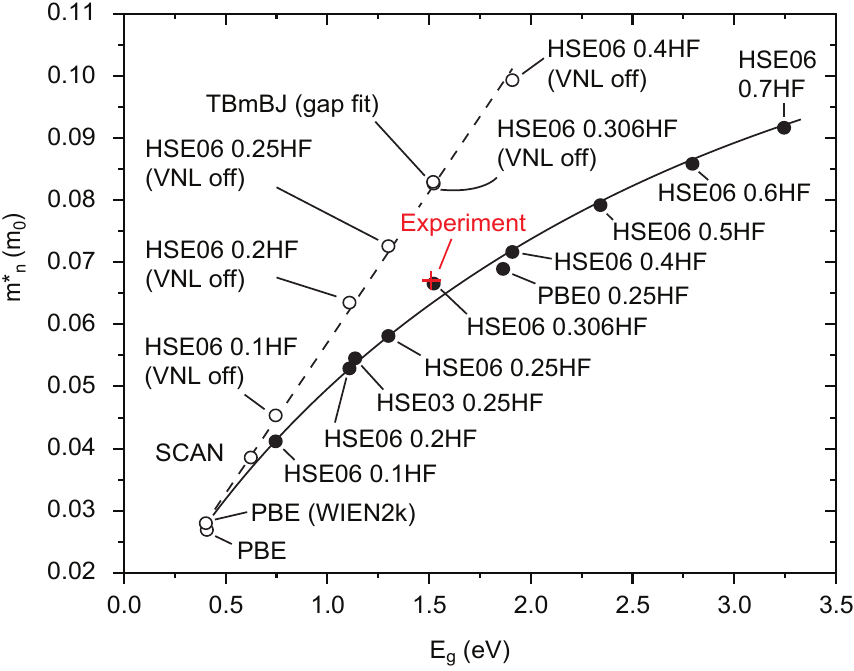}
    \caption[]{Effective mass of the conduction band of GaAs at $\Gamma$ versus the band gap for different settings of the exchange-correlation functional. There is a linear relationship between the effective mass and the band gap for all exchange-correlation functionals that do not include nonlocal exchange. Effective masses obtained from hybrid functionals considering nonlocal exchange deviate from that linear relationship. Only upon including nonlocal exchange both the experimental band gap and experimental effective mass can be reproduced correctly in the calculation. Lines are a guide to the eye.}
    \label{GaAs-mstar-Eg}
\end{figure}

Figure~\ref{GaAs-mstar-Eg} shows the effective mass of the conduction band of GaAs at $\Gamma$ versus the band gap for different settings of the exchange-correlation functional. For this graph, the electron effective mass was approximated as: 
\begin{equation}\label{Eq:mstar-GaAs}
\frac{m_{0}}{m_{\text{n}}^{*}} \approx 1 + \frac{2}{m_{0}} \sum_{\mathrm{v}} \frac{|p_{\mathrm{c} \mathrm{v}}|^2}{E_{\mathrm{c}}-E_{\mathrm{v}}}, 
\end{equation}
where the `c' and `v' indices stand for the conduction band and valence bands, respectively. In GaAs, the effective mass of the conduction band at the $\Gamma$ point is isotropic and therefore the matrix element contribution can be expressed as $2|p_{\mathrm{c} \mathrm{v}}|^2$. In the sum, we included the heavy-hole, light-hole and split-off bands as valence bands. As not all bands are included in the sum, this is an approximation. 
The matrix elements are calculated in the length gauge using VASP. 

In Fig.~\ref{GaAs-mstar-Eg} we observe a linear relationship between the effective mass and the band gap for all XC functionals that do not include nonlocal exchange. This is a consequence of the fact that $\sum{|p_{\mathrm{c} \mathrm{v}}|^2}$ changes very little, and thus the change in effective mass depends only on the change of the band gap. Effective masses obtained from hybrid functionals considering nonlocal exchange deviate from that linear relationship. They lie on a curved line that is given as guide to the eye. We also included data obtained with hybrid functionals with different screening lengths (HSE03 and PBE0) in Fig.~\ref{GaAs-mstar-Eg}. We find, that these datapoints, too, fall onto the curved line, suggesting a consistent relationship between the optical matrix element $\sum{|p_{\mathrm{c} \mathrm{v}}|^2}$ and the bandgap size. 

Only upon including nonlocal exchange both the experimental band gap and experimental effective mass can be reproduced correctly in the calculation. Figure~\ref{GaAs-mstar-Eg} thus shows the importance of including the nonlocal exchange on the HSE level for calculating accurate matrix elements and thus accurate effective masses when employing perturbation theory. The same matrix elements are also used in the calculation of dielectric properties, which explains the superior accuracy of HSE for the high-frequency dielectric constant of semiconductors and small-gap insulators~\cite{paier_dielectric_2008}.

The deviation from the linear relationship indicates that $\sum{|p_{\mathrm{c} \mathrm{v}}|^2}$ changes when the nonlocal exchange potential $V_x^{\mathrm{NL}}$ is included in the calculation of the matrix element. Thus, the matrix element is the key parameter we need to consider if we want to explain the superior accuracy of HSE effective masses, especially compared to TB-mBJ results.

\begin{figure}[t]
	\includegraphics[scale=1, angle=0]{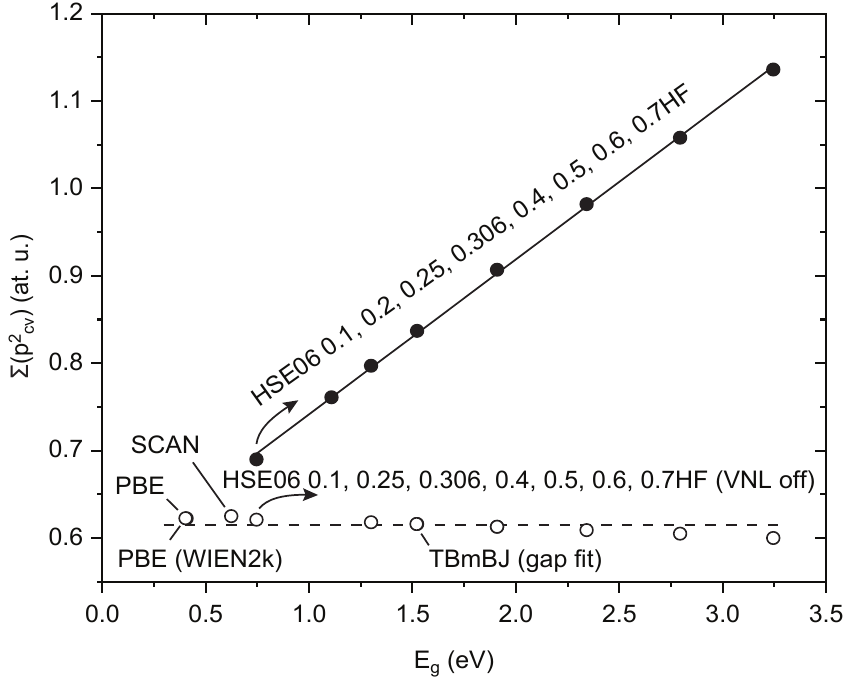}
    \caption[]{Sum of the squared matrix element $|p_{\mathrm{c} \mathrm{v}}|^2$ (see Eq.~(\ref{Eq:mstar-GaAs})) versus the band gap for different settings of the exchange-correlation potential for the conduction band effective mass of GaAs at $\Gamma$. $\sum{|p_{\mathrm{c} \mathrm{v}}|^2}$ includes contributions from the transitions between the conduction band and the heavy-hole, light-hole and split-off valence bands. $\sum{|p_{\mathrm{c} \mathrm{v}}|^2}$ is roughly a constant for all exchange-correlation functionals that do not include nonlocal exchange, irrespective of the band gap. When the nonlocal exchange is considered via the $V_x^{\mathrm{NL}}$ term in the calculation of the matrix element, the sum increases with increasing HF proportion and increasing band gap. This graph shows the importance of including the nonlocal exchange \ORrm{on} \ORadd{at (?)} the HSE level for calculating accurate optical matrix elements.}
    \label{GaAs-pcv-Eg}
\end{figure}

Figure~\ref{GaAs-pcv-Eg} shows the sum of the squared matrix elements $\sum{|p_{\mathrm{c} \mathrm{v}}|^2}$ that enter into Eq.~(\ref{Eq:mstar-GaAs}) versus the band gap for different settings of the exchange-correlation functional for the conduction band effective mass of GaAs at $\Gamma$. Again, the sum displayed on the vertical axis includes contributions from the transitions between the conduction band and the heavy-hole, light-hole and split-off valence bands. $\sum{|p_{\mathrm{c} \mathrm{v}}|^2}$ is around 0.6 atomic units for all XC functionals that do not include nonlocal effects, irrespective of the band gap. We included data obtained with the all-electron DFT code WIEN2k \cite{blaha_wien2k_2019,blaha_wien2k_2000} at the PBE level to ensure accuracy of pseudopotential calculations. We also calculated the matrix elements from the semilocal SCAN~\cite{sun_strongly_2015} functional, which gives the same matrix element as TB-mBJ and PBE in spite of the band gap being intermediate between PBE and TB-mBJ. When the nonlocal exchange potential $V_x^{\mathrm{NL}}$ is considered in the calculation of the matrix element, $\sum{|p_{\mathrm{c} \mathrm{v}}|^2}$ increases with increasing HF proportion and increasing the band gap. The change of the sum $\sum{|p_{\mathrm{c} \mathrm{v}}|^2}$ is the key that leads to the deviation of full HSE results from the linear pattern of Fig.~\ref{GaAs-mstar-Eg}. Interestingly, the increase of $\sum{|p_{\mathrm{c} \mathrm{v}}|^2}$ is strictly proportional to the increase of the HF percentage included in the functional.

\begin{figure}[h] 
	\includegraphics[scale=1, angle=0]{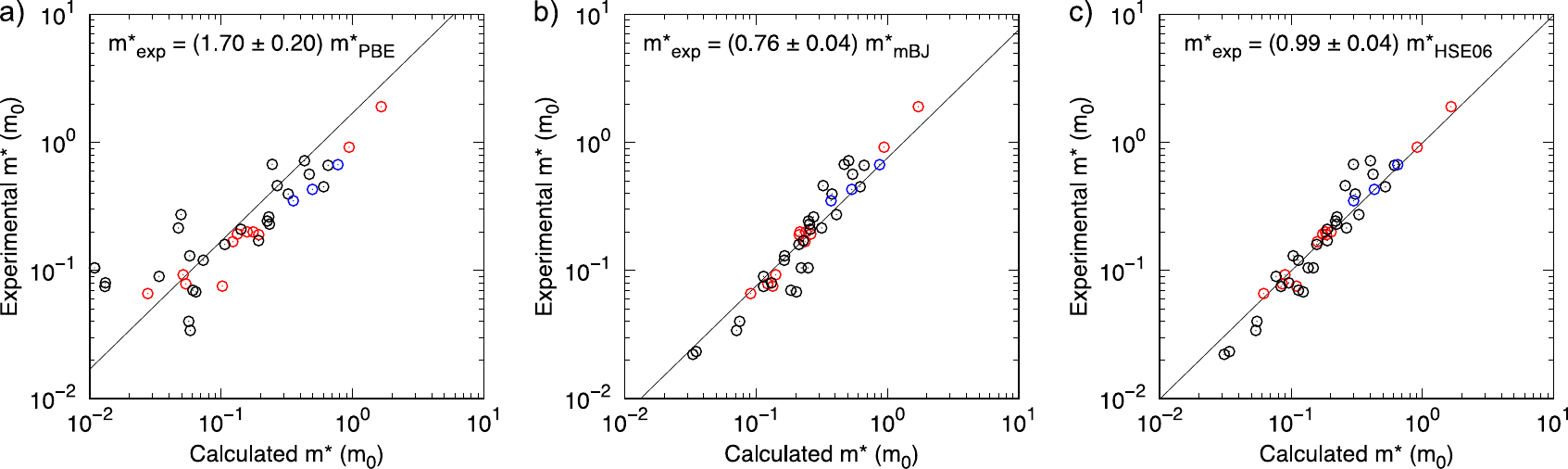}
    \caption[]{Experimental effective masses versus calculated effective masses plotted for data entries of the \texttt{mstar60} dataset included in the statistical analysis. a) PBE, b) TB-mBJ, c) HSE06. The line and coefficient obtained from linear regression are also shown. HSE06 results show the best agreement with experiment with a coefficient of $0.99 \pm 0.04$ between experimental and HSE06 effective masses. Colours represent the main contribution to the orbital composition of the band: s-orbitals (red), p-orbitals (black), and d-orbitals (blue).}
    \label{overview}
\end{figure}

Having analyzed the effect of the nonlocal exchange on the transition matrix element, we can interpret the general trends observed for effective masses in Fig.~\ref{overview} and the summary statistics of Table~\ref{stats-bulk}. Starting with TB-mBJ calculated effective masses, the clear trend for overestimation comes from the too-small absolute matrix elements. This is the same for HSE effective masses for which $V_x^{\mathrm{NL}}$ was neglected in the calculation of the matrix element. On the other hand, for PBE calculated effective masses, no clear trend is apparent. For some PBE effective masses, e.g. Si $\mathrm{m}_{\mathrm{n}, \perp}$ and $\mathrm{m}_{\mathrm{n}, \parallel}$, the agreement with experiment is surprisingly good. This is because the PBE band gap does not reproduce the experimental band gap correctly. As a result, errors in the effective mass due to an underestimated band gap and due to the underestimated matrix elements partially cancel out. This error cancellation is not systematic as seen by the high standard deviation of the relative error. Therefore, no clear trend for the error of PBE effective masses can be found. 

Transitions between bands with highly localized states (often d-states) are more affected by the exclusion of the nonlocal potential~\cite{read_calculation_1991}. Therefore, it is reasonable to test whether the error of the effective mass correlates with the orbital composition of the band of interest. To visualise that, we plotted the effective masses in Fig.~\ref{overview} where the color indicates the dominant orbital type. We can conclude that the prediction power of the functionals benchmarked in this study is not sensitive to the orbital composition of electronic states for which the effective mass was evaluated.

Finally, we look at the experimental effective masses for the transition metal dichalcogenide monolayers recorded in Table~\ref{table-1}. We observe that the bulk MoS$_2$ hole effective mass at $\Gamma$ is much lighter than the 1L MoS$_2$ effective mass. To explain this effective mass renormalization, we can consider two factors (drawing on Eq.~(\ref{Eq:mstar})): the band gap change and the change of the matrix element. The band gap at $\Gamma$ opens up from 2.08~eV in the bulk to 2.84~eV in the monolayer MoS$_2$ at the PBE level. Looking into the matrix elements, we observe that the coupling of the lower-lying conduction bands with the valence band at $\Gamma$ contributes significantly to the band dispersion. Thus, an increase of the band gap affects the strength of the contribution of these interactions to the band curvature. However, the band gap renormalization accounts for only less than half of the effective mass renormalization. This means that also the matrix elements themselves change: In 1L MoS$_2$ the sum of these matrix elements is only about half of that in bulk MoS$_2$. In other words, the oscillator strength between the valence band (VB) at $\Gamma$ and the lower-lying conduction bands is much weaker in the monolayer than in the bulk. As a result, the band dispersion of the VB at $\Gamma$ is flatter in 1L MoS$_2$ and the effective mass is larger.

\section{Conclusion}

In conclusion, we benchmarked the performance of three exchange-correlation potentials for the calculation of effective masses. Our results show that the hybrid HSE06 functional yields by far the most accurate effective masses with respect to experiment, followed by the TB-mBJ functional. We found the following ratios between experimental and calulated effective masses: $1.70 \pm 0.20$ for PBE, $0.76 \pm 0.04$ for TB-mBJ and $0.99 \pm 0.04$ for HSE06. 

We investigated the reasons behind the superior accuracy of hybrid functionals in the calculation of effective masses. After excluding bandgap effects by using fitted functionals, we focussed on the impact of the nonlocal potential on the calculation of the optical matrix element. We show that the nonlocal exchange in HSE06 enlarges the sum of the transition matrix elements which proves to be the key to the superior accuracy in the calculation of effective masses. The omission of the commutator between the nonlocal XC potentials and position when calculating optical matrix elements in HSE06 leads to serious errors (about 30\% underestimated $p_\text{cv}^2$). For the semilocal PBE functional, the errors introduced by the band gap and the transition matrix elements partially cancel out for the calculation of effective masses. For the TB-mBJ functional, PBE-like underestimated matrix elements paired with nearly experimental bandgaps lead to a consistent overestimation of effective masses. Finally, we discussed at the example of transition metal dichalcogenide bulk and monolayer materials that changes in the matrix elements are important in understanding the layer-dependent effective mass renormalization. 
In this, our analysis goes beyond the standard discussion that focusses on the interband energy difference. Our results show that changes in the matrix elements may not be ignored in the discussion of effective mass renormalization effects.

\begin{acknowledgments}
The authors are thankful to Peter Blaha (TU Vienna) and Fabien Tran (VASP Software GmbH, Vienna) for the insightful discussion about the performance of a (local) TB-mBJ potential for transition metal dichalcogenides. The authors acknowledge funding provided by the Natural Sciences and Engineering Research Council of Canada under the Discovery Grant Programs RGPIN-2020-04788. Calculations were performed using the Compute Canada infrastructure supported by the Canada Foundation for Innovation under John R. Evans Leaders Fund.
\end{acknowledgments}

\newpage

\bibliography{mstar60.bib}

\clearpage

\clearpage

\clearpage

\clearpage

\begin{table*}

\caption{Calculated and experimental effective masses $m^*$ ($m_0$). $\mathrm{m}_{\mathrm{n}}$ denote effective masses of the conduction band, $\mathrm{m}_{\mathrm{p}}$ effective masses of the valence band. `hh', `lh' and `so' stand for heavy hole, light hole and split-off band, respectively. Locations and directions in the Brillouin zone are indicated. The space group number of each material is given in parenthesis.}   
\label{table-1}
\resizebox{\textwidth}{!}{%
\begin{ruledtabular}
\begin{tabular}{llcccc}
           &                                        & experiment & PBE    & HSE06  & TB-mBJ  \\ 
\hline
Si (227)   & $\mathrm{m}_{\mathrm{n}, \perp} $  (CBM)     & 0.191      & 0.193  & 0.186  & 0.213  \\
           & $\mathrm{m}_{\mathrm{n}, \parallel}$ (CBM)   & 0.916      & 0.943  & 0.914  & 0.944  \\
           & m$_{\mathrm{p,hh}}$ ($\Gamma$) [100]         & 0.46~\cite{dexter_effective_1954}      & 0.267  & 0.259  & 0.324   \\
           & m$_{\mathrm{p,lh}}$ ($\Gamma$) [100]         & 0.171~\cite{dexter_effective_1954}     & 0.193  & 0.189  & 0.229   \\
           & m$_{\mathrm{p,so}}$ ($\Gamma$)               & 0.262      & 0.230  & 0.224  & 0.274 \\ 
\hline
GaAs (216) & m$_\mathrm{n}$  ($\Gamma$)                   & 0.066      & 0.028  & 0.062  & 0.091  \\
           & $\mathrm{m}_{\mathrm{n}, \perp} $ (X6)       & 0.23       & 0.233  & 0.223  & 0.255  \\
           & $\mathrm{m}_{\mathrm{n}, \parallel}$ (X6) $^\dagger$ \footnote{excluded from the statistical analysis}   & 1.3    & $-$0.534 & -0.812  & -0.293  \\
           & $\mathrm{m}_{\mathrm{n}, \perp} $ (L6)       & 0.075      & 0.102  & 0.110  & 0.134  \\
           & $\mathrm{m}_{\mathrm{n}, \parallel}$  (L6)   & 1.9        & 1.66  & 1.66  & 1.72  \\
           & m$_{\mathrm{p,hh}}$ ($\Gamma$)  [100]        & 0.395      & 0.324  & 0.309  & 0.379  \\
           & m$_{\mathrm{p,lh}}$ ($\Gamma$)  [100]        & 0.09       & 0.034  & 0.077  & 0.114  \\
           & m$_{\mathrm{p,so}}$  ($\Gamma$)              & 0.16       & 0.107  & 0.157  & 0.212  \\ 
\hline
GaN (186)  & $\mathrm{m}_{\mathrm{n}, \parallel}$ ($\Gamma$)    & 0.2        & 0.158  & 0.183  & 0.216    \\
           & $\mathrm{m}_{\mathrm{n}, \perp} $    ($\Gamma$)    & 0.2        & 0.175  & 0.202  & 0.239   \\ 
\hline
InP (216)  & m$_\mathrm{n}$  ($\Gamma$)                   & 0.079      & 0.054  & 0.086  & 0.123  \\
           & m$_{\mathrm{p,hh}}$  ($\Gamma$) [100]        & 0.565      & 0.469  & 0.421  & 0.543 \\
           & m$_{\mathrm{p,lh}}$  ($\Gamma$) [100]        & 0.12       & 0.073  & 0.114  & 0.164 \\
           & m$_{\mathrm{p,so}}$ ($\Gamma$)               & 0.21~\cite{rochon_photovoltaic_1975}      & 0.142  & 0.189  & 0.259  \\ 
\end{tabular}
\end{ruledtabular}}
\end{table*}

\setcounter{table}{0}
\begin{table*}
\caption{(continued)}   
\begin{ruledtabular}
\begin{tabular}{llcccc}                   
          &                                        & experiment & PBE    & HSE06   & TB-mBJ  \\ 
\hline
CdS (186)      & $\mathrm{m}_{\mathrm{n}, \perp} $ ($\Gamma$, A exciton)            & 0.192      & 0.133  & 0.174  & 0.261    \\
           & $\mathrm{m}_{\mathrm{n}, \parallel}$ ($\Gamma$, A exciton)             & 0.168      & 0.123  & 0.159  & 0.236     \\
           & $\mathrm{m}_{\mathrm{p}, \perp} $ ($\Gamma$, A exciton)                & 0.675      & 0.245  & 0.299  & 0.469     \\
           & $\mathrm{m}_{\mathrm{p}, \parallel}$ ($\Gamma$, A exciton) $^\dagger$  \footnote{excluded from the statistical analysis}            & 5          & 2.30 & 1.73  & 2.50   \\ 
\hline
CdTe (216) & m$_\mathrm{n}$   ($\Gamma$)                            & 0.093      & 0.051  & 0.90  & 0.141 \\
           & m$_{\mathrm{p,lh}}$ [100]    ($\Gamma$)                & 0.13       & 0.058  & 0.104  & 0.165 \\
           & m$_{\mathrm{p,hh}}$ [100]   ($\Gamma$)                 & 0.72       & 0.431  & 0.403  & 0.507  \\ 
\hline
PbS (225)  & $\mathrm{m}_{\mathrm{n}, \perp} $ (L)                  & 0.08     & 0.013  & 0.096  & 0.131  \\
           & $\mathrm{m}_{\mathrm{n}, \parallel}$ (L)               & 0.105    & 0.011  & 0.135  & 0.221  \\
           & $\mathrm{m}_{\mathrm{p}, \perp} $ (L)                  & 0.075    & 0.013  & 0.084  & 0.114  \\
           & $\mathrm{m}_{\mathrm{p}, \parallel}$ (L)               & 0.105    & 0.011  & 0.148  & 0.249  \\ 
\hline
PbSe (225) & $\mathrm{m}_{\mathrm{n}, \perp} $ (L)                  & 0.04     & 0.057  & 0.055  & 0.075  \\
           & $\mathrm{m}_{\mathrm{n}, \parallel}$ (L)               & 0.07     & 0.061  & 0.114  & 0.184  \\
           & $\mathrm{m}_{\mathrm{p}, \perp} $ (L)                  & 0.034    & 0.058  & 0.054  & 0.071  \\
           & $\mathrm{m}_{\mathrm{p}, \parallel}$ (L)               & 0.068    & 0.064  & 0.124  & 0.203  \\ 
\hline
PbTe (225) & $\mathrm{m}_{\mathrm{n}, \perp} $  (L)                 & 0.022    & 0.003  & 0.031  & 0.033  \\
           & $\mathrm{m}_{\mathrm{n}, \parallel}$    (L)            & 0.215    & 0.047  & 0.265  & 0.316  \\
           & $\mathrm{m}_{\mathrm{n}, \perp} $   (L)                & 0.023    & 0.003  & 0.034  & 0.035 \\
           & $\mathrm{m}_{\mathrm{p}, \parallel}$    (L)            & 0.273    & 0.050  & 0.329  & 0.410 \\ 
\hline
SiC (216)  & $\mathrm{m}_{\mathrm{n}, \parallel}$   (X)             & 0.662    & 0.652  & 0.613  & 0.665 \\
           & $\mathrm{m}_{\mathrm{n}, \perp} $      (X)             & 0.244    & 0.225  & 0.218  & 0.251 \\
           & m$_\mathrm{p}$        ($\Gamma$) [100]                 & 0.45 \footnote{specifics of the valence band and direction of the experimentally obtained effective mass are unclear}      & 0.603  & 0.523  & 0.620 \\ 
\hline
BN~\cite{andreev_influence_1994} (194)        & m$_\mathrm{p}$ ($\bar{\mathrm{K}}$, $\bar{\Gamma}$-$\bar{\mathrm{K}}$ direction) $^\dagger$  \footnote{excluded from the statistical analysis}                     & 0.49~\cite{henck_direct_2017} & 0.971  & 0.805  & 1.196 \\ 
\end{tabular}
\end{ruledtabular}
\end{table*}

\setcounter{table}{0}
\begin{table*}
\centering\caption{(continued)}  
\begin{ruledtabular}
\begin{tabular}{llcccc}
          &                                        & experiment & PBE    & HSE06   & TB-mBJ  \\ 
\hline
bP (64)    & m$_\mathrm{n}$ (Y) [010] & 1.027~\cite{narita_far-infrared_1983}         & --- & 1.15  & 1.16  \\
           & m$_\mathrm{n}$ (Y) [001]    & 0.128~\cite{narita_far-infrared_1983}      & --- & 0.122  & 0.138  \\
           & m$_\mathrm{n}$ (Y) [100]    & 0.083~\cite{narita_far-infrared_1983}      & --- & 0.056  & 0.042 \\
           & m$_\mathrm{p}$ (Y) [010]    & 0.648~\cite{narita_far-infrared_1983}      & --- & 0.668  & 0.809 \\
           & m$_\mathrm{p}$ (Y) [001]    & 0.28~\cite{narita_far-infrared_1983}       & --- & 0.276  & 0.319 \\
           & m$_\mathrm{p}$ (Y) [100]    & 0.076~\cite{narita_far-infrared_1983}      & --- & 0.053  & 0.041  \\
\hline
MoS$_2$~\cite{james_crystal_1963} (194)    & m$_\mathrm{p}$ ($\bar{\Gamma}$, $\bar{\Gamma}$-$\bar{\mathrm{K}}$ direction)  & 0.67~\cite{jin_direct_2013}      & 0.775\footnote{Perturbation theory (PT) result has an error of 14\% with respect to the band curvature fit}  & 0.649  & 0.873 \\ 
\hline
WS$_2$(194)      & m$_\mathrm{p}$ ($\bar{\mathrm{K}}$, $\bar{\Gamma}$-$\bar{\mathrm{K}}$ direction)   & 0.35~\cite{tanabe_band_2016}       & 0.354\footnote{PT result has an error of 5\% with respect to the band curvature fit} & 0.300  & 0.372 \\ 
                 & m$_\mathrm{p,VB-1}$ ($\bar{\mathrm{K}}$, $\bar{\Gamma}$-$\bar{\mathrm{K}}$ direction) & 0.43~\cite{tanabe_band_2016}       & 0.495 \footnote{PT result has an error of 7\% with respect to the band curvature fit} & 0.431  & 0.536  \\
\hline
1L MoS$_2$  & m$_\mathrm{p}$ ($\Gamma$, $\Gamma$-$\mathrm{K}$ direction)  & 2.2~\cite{jin_direct_2013,jin_substrate_2015}        & --- \footnote{Converged PT result could not be obtained. The band curvature fit gives an effective mass of 3.73 $m_0$} & 3.08 \footnote{Perturbation theory result has an error of 13\% with respect to the band curvature fit}  & ---       \\
           & m$_\mathrm{p}$ (K, $\Gamma$-$\mathrm{K}$ direction)   & 0.52~\cite{jin_substrate_2015,eknapakul_electronic_2014}    & 0.603 \footnote{PT result has an error of 15\% with respect to the band curvature fit} & 0.488 \footnote{PT result has an error of 10\% with respect to the band curvature fit} & ---       \\
           & m$_\mathrm{n}$ (K)   & 0.69~\cite{pisoni_interactions_2018,eknapakul_electronic_2014}        & 0.402 \footnote{PT result has an error of 7\% with respect to the band curvature fit}  & 0.342 \footnote{PT result has an error of 8\% with respect to the band curvature fit}   & ---       \\ 
\hline
1L MoSe$_2$ & m$_\mathrm{p}$ (K, $\Gamma$-$\mathrm{K}$ direction)  & 0.66~\cite{diaz_substrate_2017,wilson_determination_2017,kormanyos_k_2015}       & 0.672 \footnote{PT result has an error of 16\% with respect to the band curvature fit} & 0.536   & ---       \\
           & m$_\mathrm{n}$ (K)  & 0.8~\cite{larentis_large_2018}        & 0.468 \footnote{PT result has an error of 8\% with respect to the band curvature fit}  & 0.393 & ---      \\ 
\hline
1L WS$_2$   & m$_\mathrm{p}$ ($\bar{\Gamma}$, $\bar{\Gamma}$-$\bar{\mathrm{K}}$ direction) & 1.55~\cite{ulstrup_spatially_2016}       & --- \footnote{Converged PT result could not be obtained. The band curvature fit gives an effective mass of 2.93 $m_0$}  & 1.91  & ---       \\
           & m$_\mathrm{p}$ ($\bar{\mathrm{K}}$, $\bar{\Gamma}$-$\bar{\mathrm{K}}$ direction)  & 0.425~\cite{ulstrup_spatially_2016,dendzik_growth_2015}      & 0.358 \footnote{PT result has an error of 5\% with respect to the band curvature fit} & 0.305      & ---       \\
           & m$_\mathrm{p,so}$ ($\bar{\mathrm{K}}$, $\bar{\Gamma}$-$\bar{\mathrm{K}}$ direction) & 0.6~\cite{ulstrup_spatially_2016,dendzik_growth_2015}        & 0.517 \footnote{PT has an error of 8\% with respect to the band curvature fit} & 0.447     & ---       \\ 
\hline
1L WSe$_2$  & m$_\mathrm{p}$ ($\Gamma$, $\Gamma$-$\mathrm{K}$ direction)  & 4.2~\cite{wilson_determination_2017}        & --- \footnote{Converged PT result could not be obtained. The band curvature fit gives an effective mass of 4.46 $m_0$} & 2.28     & ---      
\end{tabular}
\end{ruledtabular}
\end{table*}

\clearpage

\begin{table}
\caption{Summary statistics for the error in the calculated effective mass for the bulk materials of the \texttt{mstar60} dataset. The statistics is based on 42 effective masses of 12 materials. On the right side, summary statistics of tuned functionals to reproduce experimental band gaps (BG fit) are given.}   
\label{stats-bulk}
\begin{ruledtabular}
\begin{tabular}{lccc|ccc}
           & PBE  & TB-mBJ & HSE06 & TB-mBJ & HSE06  & HSE06 $V_x^{\mathrm{NL}}$ off   \\
           &      &        &       & BG fit & BG fit & BG fit  \\
           \hline
ME (m$_\mathrm{0}$)    & $-$0.054  &  0.033  &  $-$0.028 & 0.012       & $-$0.035     & 0.034          \\
MAE (m$_\mathrm{0}$)   & 0.075     &  0.07   &  0.052  & 0.055         & 0.044        & 0.066          \\
STDE (m$_\mathrm{0}$)  & 0.100     &  0.087  &  0.095  & 0.087         & 0.090        & 0.100          \\
MRE (\%)               & $-$27     &  39     &  5.1    & 15            & $-$4.9       & 22             \\
MARE (\%)              & 38        &  45     &  21     & 21            & 11           & 26             \\
STDRE (\%)             & 40        &  49     &  29     & 21            & 16           & 22            
\end{tabular}
\end{ruledtabular}
\end{table}

\end{document}